\documentclass[12pt]{article}

\usepackage{graphicx}
\usepackage{amsmath}
\allowdisplaybreaks

\begin{document}

\baselineskip=17.5pt plus 0.2pt minus 0.1pt

\renewcommand{\theequation}{\thesection.\arabic{equation}}
\renewcommand{\thefootnote}{\fnsymbol{footnote}}
\makeatletter
\@addtoreset{equation}{section}
\def\CR{\nonumber \\}
\def\pt{\partial}
\def\be{\begin{equation}}
\def\ee{\end{equation}}
\def\bea{\begin{eqnarray}}
\def\eea{\end{eqnarray}}
\def\eq#1{(\ref{#1})}
\def\la{\langle}
\def\ra{\rangle}
\def\hyp{\hbox{-}}

\begin{titlepage}
\title{
\hfill\parbox{4cm}
{ \normalsize YITP-03-65 \\{\tt hep-th/0309035}}\\
\vspace{1cm}Non-unitary evolutions of noncommutative worlds \\ with symmetry}
\author{
Naoki {\sc Sasakura}\thanks{\tt sasakura@yukawa.kyoto-u.ac.jp}
\\[15pt]
{\it Yukawa Institute for Theoretical Physics, Kyoto University,}\\
{\it Kyoto 606-8502, Japan}}
\date{\normalsize September, 2003}
\maketitle
\thispagestyle{empty}

\begin{abstract}
\normalsize
Motivated by describing time-evolutions of noncommutative worlds, 
I discuss symmetry-preserving evolutions of noncommutative worlds of finite dimensional 
representation spaces. An interesting issue in such evolutions is  
that there can be transitions of representation spaces of symmetry, and therefore the 
evolutions are generally non-unitary or symmetry-violating.  
The central idea of this paper is that a main world evolves by emitting baby worlds 
which compensate the violations of the symmetry of the main world. 
Tracing out the states of the baby worlds,
the symmetry-preserving evolutions of the density matrices of the main world are obtained.
I give a simple example with SU(2) symmetry, which can be regarded as 
an evolving quantum two-sphere. This simple model has some attractive features resembling
our universe: it gets born from a vacuum and its entropy of 
geometric origin grows. I also discuss the evolutions in Heisenberg picture.
\end{abstract}
\end{titlepage}

\section{Introduction}
Some thought experiments with general relativity and quantum mechanics indicate that
there exist limits on the measurements of space-time observables such as 
lengths, areas, and positions 
\cite{salecker}-\cite{Adler:1999bu}.\footnote{See \cite{Yoneya:2000bt} and references 
therein for uncertainty relations in string theory.} These limits suggest an idea that
there exist some quantum natures in our space-time.
Several well-motivated algebraic
descriptions of such quantum space-times have been proposed by introducing non-commutativity 
among space-time coordinates \cite{Doplicher:1994zv}, \cite{Snyder:1946qz}-\cite{Kawamura:2003cw}. 
The study of such quantum space-times is fascinating, because it may eventually solve 
the problems on the vacuum energy \cite{Sasakura:1999xp,Thomas:2000km}
and the initial singularity of our universe.
Since our universe is expanding, a natural interesting question would be how we can formulate 
the evolutions of such quantum space-times. 

The simplest example of a quantum space would be a quantum two-sphere. 
A quantum two-sphere is defined by regarding the three SU(2) generators as the three spatial 
coordinates of a three-dimensional space. The spin of the SU(2) generators corresponds roughly
to the radius of the quantum sphere. Hence the question of time-evolutions of a 
quantum two-sphere is to find ways to relate the distinct irreducible representations of SU(2). 
Since there does not exist any SU(2)-invariant unitary maps among them, 
this question seems ambiguous if we do not impose any physical constraints.
Probably a physically reasonable evolution will be such that a perfect sphere evolves
to a perfect sphere without being deformed. Therefore the SU(2) symmetry should be kept intact. Then 
the unavoidable violation of unitarity of such evolutions must be formulated
in a physically reasonable way.

The above problem will be general for evolutions of noncommutative worlds with symmetry.
The evolutions make transitions of the representation spaces of coordinate operators,
and the question of evolutions is again to find maps among distinct representations
without breaking symmetry generated by part of the coordinate operators. I suspect that 
this problem may not be avoided even if we do not impose any symmetry generated by the 
coordinate operators, because, in physical applications, 
there will be some other important physical symmetry such as gauge symmetry and supersymmetry. 
In general situations, it will be necessary to find maps among distinct irreducible 
representations of such physical symmetry. 

A physically clear way to find such evolutions can be obtained by considering first a 
symmetry-preserving unitary process with a main world and baby worlds and then tracing out
the states of the baby worlds. Because of tracing out, we obtain evolutions of density matrices 
like in thermodynamics instead of states in representation spaces. 
The symmetry is preserved in tracing out 
the baby worlds, and therefore this is so in the evolutions of the main world.
Unitarity is lost only in the sense that there are no unitary maps among the representation spaces,
but the evolutions of the density matrices satisfy the conservation of probability: 
the traces of the density matrices are conserved under the evolutions. 
Since a density matrix defines the distributions of coordinate points, namely a kind of geometric
quantity in classical view points, we may start with a density matrix and follow its evolutions
as the evolutions of the corresponding geometry.

In the following section, I will formulate evolutions of noncommutative 
worlds along the idea in the last paragraph. 
Then in Section\,\ref{su2}, 
I will provide a simple example with SU(2) symmetry. This model describes
an evolving quantum two-sphere in the sense mentioned in the last paragraph.  
In Section\,\ref{heisenberg}, I will discuss the evolutions in Heisenberg 
picture.
The final section will be devoted to summary and discussions.
    
\section{A general formulation}
\label{general}
In this section I will formulate a symmetry-preserving evolution of a noncommutative world 
with a finite-dimensional representation space. In this paper, 
an evolution means a transition of the representation space\footnote{Other proposals of 
time-evolutions of noncommutative geometry were discussed in \cite{Connes:1994hv,Heller:1997sj}.}. 
As mentioned in the previous section, a transition cannot be 
a unitary process, when the irreducible representations of symmetry have distinct dimensions
before and after the transition. To reconcile unitarity and symmetry, let me consider a 
symmetry-preserving process that a main world splits into a main and a baby world:
\be
\label{split}
\left. |i\right> \rightarrow  {C_{i}}^{jk} \left. |j\right> \left. |k\right>,
\ee 
where the repeated indices are summed over, and $\left.|i\right>$ denotes a state in a 
representation space of the coordinates. The former state in the right-hand side represents a
state of the main world and the latter that of the baby world. 
The three-index coefficient ${C_{i}}^{jk}$ 
can be non-vanishing even when the three states are in distinct 
representations. 
The symmetry imposes that the three-index object is invariant under the symmetry transformations:
\be
\label{symmetryq}
{Q_i}^{i'} {C_{i'}}^{jk} ={C_{i}}^{j'k'} {Q_{j'}}^j {Q_{k'}}^k,
\ee
where ${Q_i}^{i'}$ denotes a symmetry operator on the states. 

The condition for the unitarity of the transition is given by 
\be
\label{unitary}
\eta_{i'i}={C^*_{i'}}^{j'k'} \eta_{j'j}\eta_{k'k} {C_{i}}^{jk}, 
\ee
where $\eta_{ij}=\left<i|j\right>$, and $*$ denotes a complex conjugate. The $\eta_{ij}$
is also assumed to be invariant under the symmetry:
\be
\label{symmetryeta}
\eta_{ij}={Q^*_i}^{i'} \eta_{i'j'} {Q_{j}}^{j'} .
\ee
It is easy to check that \eq{symmetryq}, \eq{unitary} and \eq{symmetryeta} are mutually 
consistent.

\begin{figure}[bhtp]
\begin{center}
\mbox{
\begin{minipage}{8cm}
\begin{center}
\includegraphics[width=1.65cm]{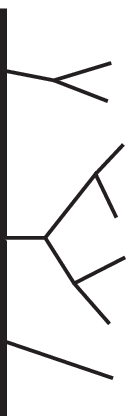}
\caption{The family tree of the main and baby worlds. The bold line represents the main 
world while the thin lines the baby worlds.}
\label{family}
\end{center}
\end{minipage}
\ \ \ \ 
\begin{minipage}{8cm}
\begin{center}
\includegraphics[width=3cm]{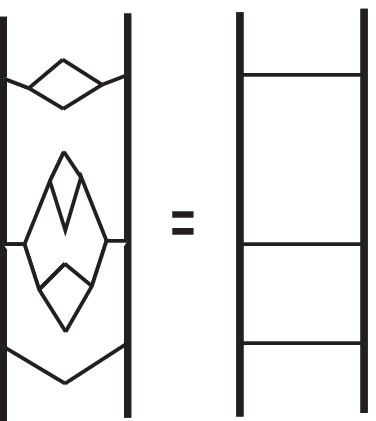}
\caption{Combining a family tree with its complex conjugate by taking the inner products for
the states of the baby worlds at the ends of the trees. 
The combined tree can be simplified to a ladder one on 
account of the unitarity of the processes of the baby worlds. }
\label{inner}
\end{center}
\end{minipage}
}
\end{center}
\end{figure}

It is natural to allow that a baby world may also split by the process \eq{split} or another
unitary process.
Then the successive applications of the process \eq{split} (or another for the baby worlds) 
lead to a family tree of worlds as in Fig.\,\ref{family}.  
The output state of the main world correlates with the states at the ends of the family tree
of the baby worlds as well as the structure of the family tree.
This is physically undesirable, since an observer of the main world cannot know the family tree 
of the baby worlds as well as the states at its ends, and therefore cannot make
predictions on the output state of the main world. 
A natural way to remove the dependence 
on the states of the baby worlds is to consider the complex conjugate of a tree and take
inner products between the corresponding states of the baby worlds (see the
left tree of Fig.\,\ref{inner}).
Then it is easy to show, by using the unitarity condition \eq{unitary} (or generally 
the unitarity of the processes of the baby worlds), that the family 
tree can be simplified to a ladder one as in Fig.\,\ref{inner}.
Thus the physically natural process of an evolution is not the one \eq{split} for the 
states in the representation space, but for the density matrices of the main world:
\be
\label{evolution}
\left. |i\right> \left< j|\right. \rightarrow {C^*_{j}}^{k'l'} \eta_{l'l} {C_{i}}^{kl} 
\left. |k \right> \left< k'|\right. .
\ee 
One can easily check that \eq{evolution} is invariant under the symmetry transformations 
${Q_i}^j$.

Another important property which a density matrix must have is the conservation of the 
probability under evolutions. Defining 
\be
{\rm Tr}[{\cal O}] \equiv \sum_{i,j} \eta^{ji} \left<i|{\cal O}|j \right>,
\ee
where $\eta^{ij}\eta_{jk}={\delta^i}_k$,
one can easily show that Tr is invariant under the evolution \eq{evolution}. 

In this section  I have obtained the evolutions of the density matrices of the main world,
which are completely independent of the evolutions of the baby worlds. 
Since there are no ways for an observer in the 
main world to detect the baby worlds, the density matrix of the main world is the only physical
reality which is predictable, while it is meaningless to talk 
about the evolutions of the states in the representation space.
This fact seems to contain an important conceptual leap 
rather than a technical limit.
Actually, density matrices have also appeared previously
in the contexts of quantum fields on curved backgrounds 
and quantum gravity \footnote{For example, see \cite{Birrell:ix} for more details.}.
Understanding the role of density matrices might be a key issue in the study of the 
dynamics of quantum geometry.  

\section{An SU(2) example}
\label{su2}
In this section I will give a simple example with SU(2) symmetry. This example turns out to be 
an evolving quantum two-sphere. 

A state in a unitary representation of SU(2) can be denoted by its spin $j$ and eigen 
value $m$ of the third component. Clebsh-Gordon 
coefficients\footnote{There are numerous elementary books containing C-G coefficients. 
For example, see \cite{Messiah}.} are SU(2)-symmetric objects which
depend on three states, take real values and satisfy
\be
\sum_{m_1=-j_1}^{j_1} \sum_{m_2=-j_2}^{j_2}
\left< j_1 m_1;j_2 m_2| J M\right> \left< j_1 m_1;j_2 m_2| J' M' \right> =\delta_{jj'}\delta_{M M'}.
\ee
Therefore one of the possible choices of the three-index coefficients ${C_i}^{jk}$ and $\eta_{ij}$
which satisfy the relations in the previous section is given by
\bea
\label{modelc}
\eta_{j_1,m_1;j_2,m_2}&=&\delta_{j_1j_2}\delta_{m_1m_2}, \cr
{C_{j,m}}^{j+1/2,m-1/2;1/2,1/2}&=&\left< j+1/2\ m-1/2;1/2\ 1/2| j\ m\right>, \cr
{C_{j,m}}^{j+1/2,m+1/2;1/2,-1/2}&=&\left< j+1/2\ m+1/2;1/2\ -1/2| j\ m\right>, \cr 
{\rm Other\ }{C_i}^{jk}&=&0.  
\eea
Note that there exist infinitely many other possible choices of the coefficients ${C_i}^{jk}$: 
I may include other spins for the baby worlds or allow the main world to shrink.
Here I have chosen the coefficients so that the main world grows monotonously with
the emissions of the smallest baby worlds with spin 1/2.
The SU(2) symmetry is not large enough to constrain a model without 
ambiguities. I hope this defect
will be absent in a more realistic model with larger symmetry.

Substituting \eq{evolution} with the SU(2) model \eq{modelc}, the evolutions of the density 
matrices are obtained as 
\bea
\label{su2evolution}
\hskip-1cm
\left. |j\ m\right>\left<j'\ m'|\right.& \rightarrow &\sqrt{\frac{(j-m+1)(j'-m'+1)}{(2j+2)(2j'+2)}}
\left. |j+1/2\ m-1/2\right>\left<j'+1/2\ m'-1/2|\right. \cr
&&+
\sqrt{\frac{(j+m+1)(j'+m'+1)}{(2j+2)(2j'+2)}}
\left. |j+1/2\ m+1/2\right>\left<j'+1/2\ m'+1/2|\right..
\eea

An interesting property of the evolutions \eq{su2evolution} is that a perfect sphere grows 
monotonously without being deformed. This is a result of the SU(2) symmetry of the evolutions.
To see that, let us define a normalized density matrix
\be
\label{densityk}
K_j=\frac1{2j+1}\sum_{m=-j}^j \left.|j\ m\right>\left<j\ m|\right..
\ee
This density matrix is invariant under the SU(2) transformation, and hence describes 
a perfect quantum two-sphere. Applying the evolutions \eq{su2evolution} to the density matrix 
\eq{densityk}, one obtains the evolution, 
\be
K_j \rightarrow K_{j+1/2}.
\ee
Therefore if one starts with the only density matrix $K_0$ of the trivial representation, 
this evolution resembles the process that a universe gets born from a vacuum and grows.
This provides a scenario for the generation of our universe.

A prominent feature of this model is that there exists entropy associated to
geometry.
The definition of the geometric entropy for a density matrix $\rho$ will be just given 
by following that of thermodynamics:
\be
S=-{\rm Tr}[ \rho \log \rho]=-{\rm Tr}[K_j \log K_j]=\log(2j+1).
\ee
At the birth of the universe $j=0$, there is no entropy $S=0$, but it grows with the 
logarithm of the number of the steps of the evolution.
There is a long-standing problem of what is the origin of the entropy of our universe.
This simple model suggests an answer: it is generated because of the non-unitary
evolutions of the geometry of the main world. It is highly possible that part of a geometric 
entropy can turn to that of matters, which is observed, 
because we expect that the degrees of freedom of geometry and matters are intertwined with 
each other in unified theory of gravity and matters.   
 
\section{Evolutions in Heisenberg picture}
\label{heisenberg}
So far I have discussed the evolutions in Schrodinger picture: what evolves is a density
matrix. Since \eq{evolution} is formally a transition rule for an operator, it can be
used to define evolutions in Heisenberg picture.
In usual quantum mechanics, Schrodinger picture and Heisenberg picture give equivalent physical results. 
The difference is just where the unitary operator of evolutions operates:
\be
(\left<f|\right.U^\dagger) {\cal O}_1 {\cal O}_2 \cdots (U \left.|i\right>)=
\left<f|\right. (U^\dagger {\cal O}_1 U)(U^\dagger {\cal O}_2 U) \cdots \left.|i\right>.
\ee
On the other hand, an evolution of a physical observable or a density matrix in the main world 
is represented schematically by 
\be
{\rm tr}_{b}[U^\dagger {\cal O} U],
\ee
where ${\rm tr}_{b}$ symbolizes tracing out baby worlds, and ${\cal O}$ is a physical
observable or a density matrix.  
Since, in general,
\be
{\rm Tr}[{\rm tr}_{b}[U \rho U^\dagger] {\cal O}_1 {\cal O}_2 \cdots ] \neq
{\rm Tr}[\rho\, {\rm tr}_{b}[U^\dagger {\cal O}_1 U] {\rm tr}_{b}[U^\dagger {\cal O}_2 U] \cdots ] ,
\ee
there will not exist the equivalence between Schrodinger and Heisenberg pictures in general.

Another interesting aspect of Heisenberg picture is that algebraic relations among
operators are not conserved in the evolutions of the main world. In usual quantum mechanics,
\be
U^\dagger {\cal O}_1 {\cal O}_2 U=(U^\dagger {\cal O}_1 U)(U^\dagger {\cal O}_2 U),
\ee
while in the evolutions of the main world
\be
{\rm tr}_b[U^\dagger {\cal O}_1 {\cal O}_2 U]\neq {\rm tr}_b[U^\dagger {\cal O}_1 U]
{\rm tr}_b[U^\dagger {\cal O}_2 U]
\ee
in general. In the usual algebraic approaches to quantum geometry, the algebraic relations 
are fixed. On the other hand, the above fact
opens a new possibility for the dynamics of noncommutative 
worlds: the algebraic relations can evolve by non-unitary operations. For example, 
one may construct a model 
which starts with large non-commutativity and approaches gradually to vanishing non-commutativity.
Such kinds of models may be interesting as the models of our universe, since large 
non-commutativity just after its birth may have observable effects on our present universe 
while there remains only negligible non-commutativity at present.

To see what actually occurs in Heisenberg picture in the SU(2) model of the previous 
section, let us define the three generators of SU(2) in the spin-$j$ representation as
\bea
\label{su2generators}
\sigma^j_3&=&\sum_{m=-j}^j m \left.|j\ m\right>\left<j\ m|\right., \cr
\sigma^{j}_{\pm}&=&\sum_{m=-j}^{j} \sqrt{j(j+1)-m(m\pm1)} \left.|j\ m\pm1\right>
\left<j\ m|\right.,
\eea
which satisfy
\bea
[\sigma^j_3, \sigma^j_{\pm}]&=&\pm \sigma^j_\pm,\cr
[\sigma^j_+, \sigma^j_-]&=&2\sigma^j_3.
\eea
Applying \eq{evolution} to \eq{su2generators}, one obtains the evolutions:
\bea
\sigma^j_3&\rightarrow& \frac{2j}{2j+2} \sigma^{j+1/2}_3,\cr
\sigma^j_{\pm}&\rightarrow& \frac{2j}{2j+2} \sigma^{j+1/2}_{\pm}.
\eea
Thus for example, if one starts with the spin-one-half generators, then after $2j-1$ steps,
one obtains
\bea
x^j_3&=&\frac{1}{j(2j+1)} \sigma^j_3,\cr
x^j_\pm&=&\frac{1}{j(2j+1)} \sigma^j_\pm.
\eea
These operators satisfy
\bea
[x^j_3, x^j_{\pm}]&=&\pm \frac{1}{j(2j+1)}\, x^j_\pm,\cr
[x^j_+, x^j_-]&=&\frac{2}{j(2j+1)}\, x^j_3.
\eea
Thus, as the evolution proceeds, the non-commutativity of the 
coordinates gradually vanishes in this SU(2) model.  

In this section, I have presented evolutions in Heisenberg picture. Since it is mathematically
inequivalent with Schrodinger picture discussed in the previous sections, 
it must be determined by some physical discussions which picture is appropriate.
In the example of SU(2), both pictures seem to provide physically reasonable results. 
I hope further study will reveal more on the two pictures and their relations.  
 
\section{Summary and discussions}
\label{discussions}
In this paper, I have discussed symmetry-preserving evolutions of noncommutative worlds
with finite-dimensional representation spaces. A non-trivial issue in such evolutions is
to find a way to reconcile unitarity and symmetry, because there can be transitions 
between distinct representations of symmetry. Using the idea of emitting baby universes,
I have described consistent evolutions in terms of density matrices. The principle behind
will be applicable to general evolutions of noncommutative worlds, and not just to
noncommutative worlds with symmetry and finite degrees of freedom.

I have also presented a simple example with SU(2) symmetry. The model describes a
monotonously growing quantum two-sphere which gets born from a vacuum.
There exists entropy of geometric origin and it grows logarithmically, starting with zero
at its birth.  These features seem to be attractive as a toy model of our universe.

I have also discussed the evolutions in Heisenberg picture. Unlike the usual quantum mechanics,
Heisenberg and Schrodinger pictures are mathematically inequivalent. In Heisenberg picture, 
the algebraic relations of noncommutative coordinates can evolve.  
Hence the evolutions in Heisenberg picture give
a potentially interesting new direction to the study of noncommutative geometry.

An important issue which was not discussed in this paper is the role of locality in evolutions.
In the models with SU(2), there are infinitely many other possible choices of the splitting 
processes of a main and baby worlds. 
Especially, it is not possible to suppress the processes of emitting 
large baby universes only by the SU(2) symmetry, although such processes seem physically 
unlikely in view of locality. Another related matter is that the discrete steps of evolutions
occur synchronously all over the quantum two-sphere in the model. This is also physically unlikely. 
This point will be physically important, because non-synchronous evolutions may generate 
some primordial fluctuations of quantum geometry. At present these problems are beyond my
reach, but might be solved by considering an algebra with large symmetry respecting locality.

Finally, it would be highly interesting to study the possibility of detecting the non-unitary 
evolutions discussed in this paper. Since each step of the evolutions is a non-unitary process, 
the non-unitarity can accumulate to be detected, if period of observation is long enough. 
Actually, some uncertainty relations of the kind that 
the uncertainty becomes larger for longer period of observation have been proposed
by several authors \cite{salecker,karolyhazy,Ng:1993jb,Amelino-Camelia:1994vs,Sasakura:1999xp}. 
The possibility of detecting this kind of uncertainty relations by gravity-wave interferometers 
was argued in \cite{Amelino-Camelia:1998ax}. I hope I can discuss these
issues in future publication.  

\vspace{.5cm}
\noindent
{\large\bf Acknowledgments}\\[.2cm]
I would like to thank the organizers of the summer school ``Mt.~Fuji 2003'' held at Shizuoka 
Prefecture, Aug.\,13-19 2003, where I thought
about the subject of this paper in its free and relaxing atmosphere.
The author was supported by the Grant-in-Aid for Scientific Research No.13135213 
from the Ministry of Education, Science, Sports and Culture of Japan.

\end{document}